\documentclass{article}
\usepackage[german, english]{babel}
\usepackage[a4paper, left=2.5cm, right=2.5cm, top=3.5cm, bottom=3cm]{geometry}
\usepackage{amssymb}
\usepackage{amsmath}
\usepackage{amscd}
\usepackage{latexsym}
\usepackage{graphicx}
\usepackage{ifthen}
\usepackage{epsfig}
\usepackage{setspace}
\usepackage{subcaption}
\usepackage{hyperref}
\usepackage{dsfont}
\usepackage{mathrsfs}
\usepackage{color}
\hypersetup{
 colorlinks=true,
 linkcolor=blue,
 citecolor=magenta,
}

\catcode`@=11
\renewcommand{\section}{\@startsection{section}{1}{0pt}{\medskipamount}
{\medskipamount}{\large\bf}}
\numberwithin{equation}{section}
\catcode`@=12

\newcommand{\C}{\mathds C}
\newcommand{\N}{\mathds N}
\newcommand{\R}{\mathds R}

\newcommand{\Acal}{{\cal A}}
\newcommand{\Fcal}{{\cal F}}
\newcommand{\Ecal}{{\cal E}}
\newcommand{\Bcal}{{\cal B}}
\newcommand{\Dcal}{{\cal D}}
\newcommand{\Pcal}{{\cal P}}
\newcommand{\Kcal}{{\cal K}}
\newcommand{\Mcal}{{\cal M}}
\newcommand{\Jcal}{{\cal J}}
\newcommand{\tY}{{\widetilde{Y}}}

\def\im{\mathrm{i}}
\def\ep{\mathrm{e}}
\def\pa{\mbox{$\partial$}}
\def\diff{\mathrm{d}}

\def\sfrac#1#2{{\textstyle\frac{#1}{#2}}}
\def\]{\right]}
\def\[{\left[}
\def\){\right)}
\def\({\left(}
\def\>{\rangle}
\def\<{\langle}
\def\+{\dagger}
\def\we{{\wedge}}
\def\={\ =\ }
\def\und{\quad\textrm{and}\quad}
\def\with{\quad\textrm{with}\quad}
\def\for{\quad\textrm{for}\quad}
\def\3j#1#2#3#4#5#6{\begin{pmatrix} #1&#2&#3\\#4&#5&#6 \end{pmatrix}}
\def\s3j#1#2{\begin{pmatrix} #1\\#2 \end{pmatrix}}

\begin{document}

\title{\bf\huge On rational electromagnetic fields}
\date{~}

\author{\phantom{.}\\[12pt]
{\Large Olaf Lechtenfeld}$^{*\dagger}$ \
and \ {\Large Kaushlendra Kumar}$^{\dagger}$
\\[24pt]
$^{\dagger}${Institut f\"ur Theoretische Physik}\\ 
{Leibniz Universit\"at Hannover} \\ 
{Appelstra{\ss}e 2, 30167 Hannover, Germany}
\\[24pt]
$^{*}${Riemann Center for Geometry and Physics}\\ 
{Leibniz Universit\"at Hannover} \\ 
{Appelstra{\ss}e 2, 30167 Hannover, Germany}
\\[12pt] 
} 

\clearpage
\maketitle
\thispagestyle{empty}

\begin{abstract}
\noindent\large
We employ a recently developed method for constructing rational electromagnetic field configurations 
in Minkowski space to investigate several properties of these source-free finite-action Maxwell (``knot'') solutions.
The construction takes place on the Penrose diagram but uses features of de Sitter space, in particular its isometry group.
This admits a classification of all knot solutions in terms of $S^3$ harmonics, labelled by a spin~$2j\in\N_0$,
which in fact provides a complete ``knot basis'' of finite-action Maxwell fields.
We display a $j{=}1$ example, compute the energy for arbitrary spin-$j$ configurations, 
derive a linear relation between spin and helicity and characterize the subspace of null fields.
Finally, we present an expression for the electromagnetic flux at null infinity and demonstrate its equality with the total energy.
\end{abstract}

\newpage
\setcounter{page}{1} 

\section{Introduction and summary}

\noindent
Electromagnetic knot configurations in Minkowski space were discovered in 1989 by Ra\~nada~\cite{elknots} and have been an active 
field of research ever since (for a review, see~\cite{knotreview}). Their electric and magnetic fields are rational functions 
of the Cartesian Minkowski coordinates, and their field lines exhibit nontivial topology characterized by the conserved helicity.
Several methods for constructing such source-free Maxwell solutions have been developed, employing the Hopf map, Penrose twistors,
complex Euler potentials (for null fields) or special conformal transformations.

In a recent paper~\cite{lechtenfeld-zhilin} co-authored by one of us, a further method for building rational (knot) solutions 
has been found. It is based on a correspondence of Maxwell solutions on Minkowski space~$\R^{1,3}$ and on de Sitter space~dS$_4$,
thanks to the conformal equivalence between (part of) these spaces and the conformal invariance of four-dimensional gauge theory.\footnote{
This extends to Yang-Mills theory. In fact, the original motivation arose from investigating non-Abelian gauge theory on de Sitter space
\cite{lechtenfeld-popov, lechtenfeld-popov2}.}{\footnote{
Incidentally, conformal transformations of the domain space with complex parameters have also been employed 
to generate knot solutions \cite{irvine-bouwmeester}, 
but this strategy is not related to our work in an obvious way.}
The O(1,4) isometry of de Sitter space (with three-spheres as equal-time slices) suggested 
an O(4)~covariant treatment of Maxwell theory on dS$_4$, which resulted in a complete basis of vacuum electromagnetic fields, 
labelled by the weights of spin~$(j,j)$ irreps of the $so(4)\simeq su(2)_L\oplus su(2)_R$ spatial isometry algebra.
The isomorphism $S^3\simeq\textrm{SU}(2)$ further allowed one to impose left invariance (under the group multiplication).
Mapping those basis solutions to Minkowski space provided a straightforward algorithm for generating a full basis of finite-action
rational Maxwell solutions -- electromagnetic knots. The method was then illustrated on a couple of examples, which demonstrated
that very complex Minkowski-space configurations are generated from rather simple de Sitter-space expressions.

To further test the ``de Sitter method'' and to establish its usefulness, it is warranted to investigate various properties
of electromagnetic knots from this new perspective and to learn how well this method does in obtaining them.
This is the main purpose of the paper. To its end, we (a)~revisit and streamline the construction and apply it to
an explicit $j{=}1$ example,
(b)~compute the conserved energy and helicity, (c)~characterize the subspace of null fields 
($\vec{E}^2=\vec{B}^2$ and $\vec{E}\cdot\vec{B}=0$) and (d)~compute the energy flux radiated to infinity.
We also comment on the topological structure of the electric and magnetic field lines and display the spatial distribution
of the energy density for exemplary configurations with $j{=}\sfrac12$ and $j{=}1$, beyond the known $j{=}0$ case of the
Ra\~nada--Hopf knot.
In all cases, pulling Minkowski-space quantities back to de Sitter space brought on substantial simplifications.
We conclude that this novel approach to electromagnetic knots is a powerful tool, both conceptually and computationally.
In order to better understand the physical properties of these knot configurations (also for higher~$j$) one might
want to probe them with charged test particles, classically or quantum mechanically. Another future project is the
backreaction of such sources on the knot configuration.

The paper is organized as follows.
In Section~2 we introduce the calculational tools on de Sitter space needed for our construction, including the $S^3$~harmonics
(for background material, see e.g.~\cite{Jantzen, Higuchi, Lachieze-Rey, Lindblom}).
We find it convenient to pass from de Sitter space to a conformally related Lorentzian cylinder over~$S^3$.
Section~3 reviews the solution of the source-free Maxwell equations on de Sitter space and gives a complete characterization 
in terms of $so(4)$ irreps of spin~$j$ and their weights.
The conformal map to Minkowski space provides the Penrose-diagram representation of the latter. 
This is the subject of Section~4, which provides the recipe for computing the rational field configurations also known as electromagnetic knots,
and illustrates it with a $j{=}1$ example.
We briefly analyze manifest and hidden symmetries on de Sitter and on Minkowski space in Section~5.
Section~6 evaluates the energy density and the helicity density, which turn out to be linearly related for a given spin~$j$,
and includes comments on the field-line topology.
The interesting subclass of null fields is studied in Section~7, where their moduli space is described as 
a complete-intersection projective variety of complex dimension~$2j{+}1$.
Field lines and energy densities for two examples are depicted.
In Section~8 we investigate the electromagnetic flux across future null infinity and show it to coincide with the field energy.

\section{Calculus on de Sitter space}

\noindent
Four-dimensional de Sitter space $\diff\mathrm{S}_4$ is an embedding of a one-sheeted hyperboloid 
in five-dimensional Minkowski space $\R^{1,4}\ni(q_0^{},q_{_A})$ with $A=1,2,3,4$ given by
\begin{equation}\label{dS}
   -q_0^{\ 2} + q_1^{\ 2} + q_2^{\ 2} + q_3^{\ 2} + q_4^{\ 2} \= \ell^2\ .
\end{equation}\label{flat_metric}
Here, the de Sitter radius~$\ell$ provides a scale, and the flat Minkowski metric 
\begin{equation}
   \diff s_5^{\ 2} \= -\diff q_0^{\ 2} + \diff q_1^{\ 2} + \diff q_2^{\ 2} + \diff q_3^{\ 2} +\diff q_4^{\ 2}
\end{equation}
induces a metric $\diff s^2$ on dS$_4$.
On this hyperboloid we choose the following intrinsic coordinates,
\begin{equation}\label{transf1}
   q_0^{} \= -\ell\,\cot\tau \und q_{_A} = \ell\,\omega_{_A} \csc\tau \quad\with \tau\in {\cal I}\equiv(0,\pi) \und A = 1,\ldots,4\ ,
\end{equation}
where the $\omega_{_A}$ subject to $\omega_{_A}\omega_{_A}=1$ embed a unit three-sphere~$S^3\ni(\chi,\theta,\phi)$ into~$\R^4$ via
\begin{equation}\label{4vector}
   \omega_1 \= \sin\chi\,\sin\theta\,\cos\phi\ ,\quad \omega_2 \= \sin\chi\,\sin\theta\,\sin\phi\ ,\quad \omega_3 \= \sin\chi\,\cos\theta\ ,\quad \omega_4 \= \cos\chi\ ,
\end{equation}
with $0\leq \chi,\theta \leq \pi$ and $0\leq\phi\leq2\pi$.
With this hyperspherical parametrization, dS$_4$ obviously is diffeomorphic to a Lorentzian cylinder ${\cal I}\times S^3$. 
More importantly, the natural cylinder metric $\diff s^2_{\textrm{cyl}}$ is actually conformal to the induced de Sitter metric,
\begin{equation} \label{dSmetric}
   \diff s^2 \ \equiv\ \diff s_5^{\ 2}\big|_{\textrm{dS}_4} \= 
   \frac{\ell^2}{\sin^2\!\tau} \left( -\diff\tau^2 + \diff\omega_{_A} \diff\omega_{_A} \bigl|_{\R^4} \right) \=
   \frac{\ell^2}{\sin^2\!\tau} \left( -\diff\tau^2 + \diff\Omega_3^{\ 2} \right) \=
   \frac{\ell^2}{\sin^2\!\tau} \,\diff s^2_{\textrm{cyl}}\ ,
\end{equation}
where $\diff\Omega_3^{\ 2}=\diff\chi^2+\sin^2\!\chi\,(\diff\theta^2+\sin^2\!\theta\,\diff\phi^2)$ denotes the round three-sphere metric.

Because vacuum electrodynamics is conformally invariant, 
Maxwell's equations on a de Sitter vacuum may equally well be solved on the cylinder~${\cal I}\times S^3$.
This is what we shall do in the following section.
To this end, we shall need a basis of one-forms defined by
\begin{equation}
e^\tau = \diff\tau \quad\und\quad
e^a \=  - \eta^a_{\ _{BC}} \, \omega_{_B}\, \diff \omega_{_C}
\end{equation}
using the self-dual `t Hooft symbol $\eta^a_{\ _{BC}}$ for $a=1,2,3$ and $B,C=1,\ldots,4$ with non-vanishing components
\begin{equation}
\eta^a_{\ bc} = \varepsilon^a_{\ bc} \quad\und\quad \eta^a_{\ b4} = -\eta^a_{\ 4b} = \delta^a_{\ b}\ .
\end{equation}
Here, we have taken advantage of the fact that
\begin{equation}
S^3 \simeq \textrm{SU}(2) \quad\und\quad so(4) \simeq su(2)_L\oplus su(2)_R
\end{equation}
by choosing the basis one-forms $e^a$ to be invariant under the dragging induced by the left SU(2) multiplication. 
They obey the useful identities
\begin{equation}
   \diff e^a + \varepsilon^a_{\ bc}\,e^b\wedge e^c \=0 \quad\und\quad e^a e^a \= \diff\Omega_3^{\ 2}\ .
\end{equation} 
Alternatively, one may obtain this basis through the left Cartan one-form 
\begin{equation}
\Omega_{_L}(g) := g^{-1}\, \diff g \= e^a\,T_a \quad\with T_a = -\sfrac{\im}{2} \sigma_a
\end{equation}
where the SU(2) generators~$T_a$ are given by the Pauli matrices~$\sigma_a$, and the identification map
\begin{equation}
g: \ S^3 \rightarrow \textrm{SU}(2) \qquad\textrm{via}\quad
(\omega_{_A}) \mapsto -\im \begin{pmatrix} \beta & \alpha^* \\ \alpha & -\beta^* \end{pmatrix}
\quad\with \alpha=\omega_1{+}\im\omega_2 \und \beta=\omega_3{+}\im\omega_4\ ,
\end{equation}
sends the $S^3$ north pole $(0,0,0,1)$ to the group identity $\mathds{1}_2$. 

Dual to the left-invariant one-form basis $\{e^a\}$ there exist left-invariant vector fields
\begin{equation}\label{vecfields}
   R_a \= -\eta^a_{_{\ BC}}\,\omega_{_B}\,\frac{\partial}{\partial \omega_{_C}} \qquad\Rightarrow\qquad
   \[ R_a , R_b \] = 2\,\varepsilon_{ab}^{\ \ c}\,R_c
\end{equation}
generating the right multiplication on SU(2) (hence the notation) whose algebra is denoted by $su(2)_R$.
The $su(2)_L$ half of the $so(4)$ isometry of~$S^3$ is provided by the right-invariant vector fields
\begin{equation}
    L_a \= -\tilde{\eta}^a_{_{\ BC}}\,\omega_{_B}\,\frac{\partial}{\partial\omega_{_C}} \qquad\Rightarrow\qquad
    \[ L_a , L_b \] = 2\,\varepsilon_{ab}^{\ \ c}\,L_c
\end{equation}
belonging to the left multiplication on the group manifold and constructed from the anti-self-dual `t Hooft symbol
$\tilde{\eta}^a_{_{\ BC}}$, which is obtained from $\eta^a_{\ _{BC}}$ by flipping the sign of the `4'~components.
The differential of a function~$f$ on~${\cal I}\times S^3$ is conveniently taken as
\begin{equation}
\diff f \= \diff\tau\,\pa_\tau f\,+\,e^a R_a f \= \diff\tau\,\pa_\tau f\,+\,\tilde{e}^a L_a f\ ,
\end{equation}
where $\{\tilde{e}^a\}$ would be a basis of right-invariant one-forms dual to $\{L_a\}$.

Functions on $S^3$ can be expanded in a basis of harmonics~$Y_j(\chi,\theta,\phi)$ with $2j\in\N_0$, 
which are eigenfunctions of the scalar Laplacian,\footnote{
The SO(4) spin of these functions is actually $2j$, but we label them with half their spin, for reasons to be clear below.}
\begin{equation}
-\mathop{}\!\mathbin\bigtriangleup_3 Y_j \= 2j(2j{+}2)\,Y_j \= 4j(j{+}1)\,Y_j \=
-\sfrac12(L^2+R^2)\,Y_j \= -\sfrac14(\Dcal^2+\Pcal^2)\,Y_j\ ,
\end{equation}
where $L^2=L_a L_a$ and $R^2=R_a R_a$ are (minus four times) the Casimirs of $su(2)_L$ and $su(2)_R$, respectively,
\begin{equation}
-\sfrac14 L^2\,Y_j \= -\sfrac14 R^2\,Y_j \= -\sfrac14\mathop{}\!\mathbin\bigtriangleup_3 Y_j \= j(j{+}1)\,Y_j\ .
\end{equation}
We have also introduced $\Dcal^2=\Dcal_a \Dcal_a$ and $\Pcal^2=\Pcal_a \Pcal_a$ with
\begin{equation} \label{DPdef}
\Dcal_a \= L_a+R_a \= -2\,\varepsilon_a^{\ bc}\,\omega_b\,\pa_c \quad\und\quad 
\Pcal_a \= L_a-R_a \= 2\,\omega_{[a}\,\pa_{4]}  \quad\with \pa_{_A}\equiv\sfrac{\pa}{\pa\omega_A}
\end{equation}
so that
\begin{equation}
\[ \Dcal_a , \Dcal_b \] \= 2\,\varepsilon_{ab}^{\ \ c}\,\Dcal_c \ ,\qquad
\[ \Dcal_a , \Pcal_b \] \= 2\,\varepsilon_{ab}^{\ \ c}\,\Pcal_c \ ,\qquad
\[ \Pcal_a , \Pcal_b \] \= 2\,\varepsilon_{ab}^{\ \ c}\,\Dcal_c \ .
\end{equation}
Hence, $\{\Dcal_a\}$ spans the diagonal subalgebra~$su(2)_D\subset so(4)$, which generates the stabilizer subgroup~SO(3) 
in the coset representation~$S^3\simeq\textrm{SO}(4)/\textrm{SO}(3)$.
Therefore, $\Dcal^2$ is (minus four times) the Casimir of~$su(2)_D$, with eigenvalues $l(l{+}1)$ for $l=0,1,\ldots,2j$,
and $\sfrac14 \Dcal^2=\mathop{}\!\mathbin\bigtriangleup_2$ is the scalar Laplacian on the $S^2$ slices traced out in~$S^3$ by the SO(3)$_D$ action.

To further characterize a complete basis of $S^3$ harmonics, there are two natural options,
corresponding to two different complete choices of mutually commuting operators to be diagonalized.
First, the left-right (or toroidal) harmonics~$Y_{j;m,n}$ are eigenfunctions of $L^2=R^2$, $L_3$ and~$R_3$,
\begin{equation} \label{Y-action}
\sfrac{\im}{2}\,L_3\,Y_{j;m,n} \= m\,Y_{j;m,n} \quad\und\quad
\sfrac{\im}{2}\,R_3\,Y_{j;m,n} \= n\,Y_{j;m,n} \ ,
\end{equation}
and hence the corresponding ladder operators 
\begin{equation}
L_\pm \= (L_1\pm\im L_2)/\sqrt{2} \quad\und\quad R_\pm \= (R_1\pm\im R_2)/\sqrt{2}
\end{equation}
act as
\begin{equation}
\sfrac{\im}{2}\,L_\pm\,Y_{j;m,n} \= \sqrt{(j{\mp}m)(j{\pm}m{+}1)/2}\,Y_{j;m\pm1,n} \quad\!\und\!\quad
\sfrac{\im}{2}\,R_\pm\,Y_{j;m,n} \= \sqrt{(j{\mp}n)(j{\pm}n{+}1)/2}\,Y_{j;m,n\pm1}\ .
\end{equation}
Second, the adjoint (or hyperspherical) harmonics~$\tY_{j;l,M}$ are eigenfunctions of $L^2=R^2$, $\Dcal^2$ and~$\Dcal_3$,
\begin{equation} 
-\sfrac{1}{4}\,\Dcal^2\,\tY_{j;l,M} \= l(l{+}1)\,\tY_{j;l,M} \quad\und\quad
\sfrac{\im}{2}\,\Dcal_3\,\tY_{j;l,M} \= M\,\tY_{j;l,M} \ ,
\end{equation}
with the ladder-operator actions~\cite{wybourne}
\begin{equation}
\begin{aligned}
\sfrac{\im}{2}\,\Dcal_\pm\,\tY_{j;l,M} &\= \sqrt{(l{\mp}M)(l{\pm}M{+}1)/2}\,\tY_{j;l,M\pm1} \ ,\\[4pt]
\sfrac{\im}{2}\,\Pcal_\pm\,\tY_{j;l,M} &\= 
\mp\sqrt{(l{\mp}M{-}1)(l{\mp}M)/2}\,c_{j,l}\,\tY_{j;l-1,M\pm1} \ 
\pm\ \sqrt{(l{\pm}M{+}1)(l{\pm}M{+}2)/2}\,c_{j,l+1}\,\tY_{j;l+1,M\pm1} \ ,\\[4pt]
\sfrac{\im}{2}\,\Pcal_3\,\tY_{j;l,M} &\= 
\sqrt{l^2{-}M^2}\,c_{j,l}\,\tY_{j;l-1,M} \ +\ 
\sqrt{(l{+}1)^2{-}M^2}\,c_{j,l+1}\,\tY_{j;l+1,M} \ ,
\end{aligned}
\end{equation}
where
\begin{equation}
c_{j,l} \= \sqrt{\bigl((2j{+}1)^2-l^2\bigr)/\bigl((2l{-}1)(2l{+}1)\bigr)}\ .
\end{equation}
In this case, there exists a recursive construction for harmonics on $S^{k+1}$ from those on $S^k$,
\begin{equation}\label{split-Y}
\tY_{j;l,M}(\chi,\theta,\phi) \= R_{j,l}(\chi)\,Y_{l,M}(\theta,\phi) \with
R_{j,l}(\chi) \= \im^{2j+l}\,\sqrt{\sfrac{2j+1}{\sin\chi}\sfrac{(2j+l+1)!}{(2j-l)!}}\, P_{2j+\frac12}^{-l-\frac12}(\cos\chi)\ ,
\end{equation}
where $Y_{l,M}$ are the standard $S^2$ spherical harmonics and $P_a^b$ denote the associated Legendre polynomials of the first kind.\footnote{
With fractional indices, it is rather a Gegenbauer polynomial, but also a hypergeometric function (see eq.~(2.8) of~\cite{Higuchi}).} 
The two bases of harmonics are related by the standard Clebsch-Gordan series for the angular momentum addition
$j\otimes j = 0\oplus 1\oplus\ldots\oplus 2j$,
\begin{equation}\label{Y-new}
Y_{j;m,n} \= \sum\limits_{l=0}^{2j}\sum\limits_{M=-l}^{l}\,C_{m,n}^{l,M}\,\tY_{j;l,M}\ ,
\qquad\textrm{with}\qquad C_{m,n}^{l,M} = \<2j;l,M|j,m;j,n\>
\end{equation}
being the Clebsch-Gordan coefficients enforcing $m{+}n{=}M$ and $l\in\{0,1,\ldots,2j\}$.

\section{Solving vacuum Maxwell equations on de Sitter space}

\noindent
The Maxwell gauge potential is a real-valued one-form on ${\cal I}\times S^3$,
\begin{equation} \label{Acalfull}
\Acal \= \Acal_\tau(\tau,\omega)\,e^\tau\ +\ \sum_{a=1}^3 \Acal_a(\tau,\omega)\,e^a 
\quad\with \omega\equiv\{\omega_{_A}\}\ .
\end{equation}
Imposing the Coulomb gauge
\begin{equation}
   \Acal_\tau(\tau,\omega) = 0\quad\und\quad R_a\,\Acal_a (\tau,\omega) = 0
\end{equation} 
simplifies the field strength to
\begin{equation}\label{2-form}
  \Fcal \= \diff\Acal \= 
  \pa_\tau\Acal_a\,e^\tau\we e^a + \bigl(\sfrac12 R_{[b}\Acal_{c]}-\Acal_a\,\varepsilon^a_{\ bc} \bigr)\,e^b\we e^c\ ,
\end{equation}
and the vacuum Maxwell equations of motion $\diff*\Fcal=0$ to
\begin{equation} \label{Maxwell}
  \pa_\tau^2\Acal_a \= (R^2-4)\,\Acal_a + 2\,\varepsilon_{abc}R_b\Acal_c\ .
\end{equation}
These are coupled linear wave equations for $S^3$.

As was shown in \cite{lechtenfeld-zhilin}, the general solution of \eqref{Maxwell} decomposes into spin-$j$ representations of~$so(4)$,
\begin{equation}\label{Asep}
\Acal_a(\tau,\omega) \=  
\Bigl\{ \sum_{2j=0}^\infty X_{a\ \textrm{I}}^j(\omega) \ \ep^{2(j+1)\im\tau} \ +\ \textrm{c.c.} \Bigr\} \ +\ 
\Bigl\{ \sum_{2j=2}^\infty X_{a\ \textrm{II}}^j(\omega) \ \ep^{2j\,\im\tau} \ +\ \textrm{c.c.} \Bigr\}\ ,
\end{equation}
consisting of a type-I contribution with frequency $\Omega_j{=}2(j{+}1)$ and a type-II contribution with frequency~$\Omega_j{=}2\,j$.
We reorganize the complex angular functions~$X_a^j$ as
\begin{equation}\label{XZ}
X_1^j=\sfrac{1}{\sqrt{2}} \bigl(Z_+^j + Z_-^j \bigr)\ ,\qquad 
X_2^j=\sfrac{\im}{\sqrt{2}} \bigl( Z_-^j - Z_+^j \bigr)\ ,\qquad 
X_3^j= Z_3^j
\end{equation}
for both types and expand the functions~$Z^j_\pm$ and $Z^j_3$ into spin-$j$ basis solutions (for $*\in\{+,3,-\}$),
\begin{equation}\label{basisZ}
Z_{*\ \textrm{I}}^j(\omega) \= \sum_{m=-j}^j \sum_{n=-j-1}^{j+1} \lambda_{j;m,n}^{\textrm{I}}\ Z_{*\ \textrm{I}}^{j;m,n}(\omega)  \quad\und\quad
Z_{*\ \textrm{II}}^j(\omega) \= \sum_{m=-j}^j \sum_{n=-j+1}^{j-1} \lambda_{j;m,n}^{\textrm{II}}\ Z_{*\ \textrm{II}}^{j;m,n}(\omega)\ ,
\end{equation}
with $(2j{+}1)(2j{+}3)$ arbitrary complex coefficients $\lambda^{\textrm{I}}_{j;m,n}$
and $(2j{+}1)(2j{-}1)$ coefficients $\lambda^{\textrm{II}}_{j;m,n}$.
(Note that type-II solutions are absent for $j{=}0$ and $j{=}\sfrac12$.)
The complex angular basis functions~$Z_*^{j;m,n}$ take the following form:
\begin{itemize}
\addtolength{\itemsep}{-4pt}
\item type I : \quad
$j{\geq}0\ ,\quad m = -j,\ldots,+j\ ,\quad n = -j{-}1,\ldots,j{+}1\ ,\quad \textrm{frequency}\ \Omega_j=2(j{+}1)\ ,$
\\
\begin{equation} \label{type1}
\begin{aligned}
Z_{+\ \textrm{I}}^{j;m,n}(\omega) &\= \sqrt{(j{-}n)(j{-}n{+}1)/2} \ \ Y_{j;m,n+1}(\omega) \ ,\\
Z_{3\ \textrm{I}}^{j;m,n}(\omega)\,&\= \sqrt{(j{+}1)^2-n^2} \ \ Y_{j;m,n}(\omega) \ ,\\
Z_{-\ \textrm{I}}^{j;m,n}(\omega)   &\= -\sqrt{(j{+}n)(j{+}n{+}1)/2} \ \ Y_{j;m,n-1}(\omega) \ .
\end{aligned}
\end{equation}
\item type II :\quad
$j{\geq}1\ ,\quad m = -j,\ldots,+j\ ,\quad n = -j{+}1,\ldots,j{-}1\ ,\quad \textrm{frequency}\ \Omega_j=2\,j\ ,$
\\
\begin{equation} \label{type2}
\begin{aligned}
Z_{+\ \textrm{II}}^{j;m,n}(\omega)  &\= -\sqrt{(j{+}n)(j{+}n{+}1)/2} \ \ Y_{j;m,n+1}(\omega)\ ,\\
Z_{3\ \textrm{II}}^{j;m,n}(\omega)\,&\= \sqrt{j^2-n^2} \ \ Y_{j;m,n}(\omega) \ ,\\
Z_{-\ \textrm{II}}^{j;m\,n}(\omega)   &\= \sqrt{(j{-}n)(j{-}n{+}1)/2} \ \ Y_{j;m,n-1}(\omega) \ .
\end{aligned}
\end{equation}
\end{itemize}
Inserting \eqref{type1} and \eqref{type2} into \eqref{basisZ} and the result into~\eqref{XZ} provides a harmonic expansion
\begin{equation}\label{XY}
X_a^j(\omega) \= \sum_{m=-j}^j \sum_{n=-j}^j X_a^{j;m,n}\ Y_{j;m,n}(\omega)
\end{equation}
for both types of angular functions in~\eqref{Asep}. 
(Note the different range of~$n$ for $X_a^{j;m,n}$ and $Z_*^{j;m,n}$; they are not easily related as $X_a^j$ and $Z_*^j$ are in~\eqref{XZ}.)

It is useful for later purposes to introduce here the ``sphere-frame" electric and magnetic fields,
\begin{equation}
    \Fcal \= \Ecal_a\,e^a \we e^\tau + \sfrac12\,\Bcal_a\,\varepsilon^a_{\ bc}\,e^b\we e^c\ .
\end{equation}
For a fixed type (I or II) and spin~$j$, we may eliminate $R_{[b}A_{c]}$ in \eqref{2-form} 
by using~\eqref{Maxwell} and employ
\begin{equation}
\pa_\tau^2 \Acal^{(j)} = -\Omega_j^2\,\Acal^{(j)} \quad\und\quad
R^2\,\Acal^{(j)} \= -4j(j{+}1)\,\Acal^{(j)}
\end{equation}
to obtain
\begin{equation}\label{cur-field}
\Ecal_a^{(j)} \= -\pa_\tau \Acal_a^{(j)} \quad\und\quad \Bcal_a^{(j)} \= \mp\Omega_j\,\Acal_a^{(j)}\ ,
\end{equation}
where the upper sign pertains to type~I and the lower one to type~II.
We note in passing that, due to the compactness of the Lorentzian cylinder, the sphere-frame energy and action are always finite.

Due to the linearity of Maxwell theory, the overall scale of any solution is arbitrary.
Furthermore, the parity transformation $L\leftrightarrow R$ and $m\leftrightarrow n$ 
interchanges a spin-$j$ solution of type~I with a spin-$(j{+}1)$ solution of type~II.
Finally, electromagnetic duality at fixed $j$ is realized by shifting $\Omega_j\tau$ by $\sfrac{\pi}{2}$ for type~I
or by $-\sfrac{\pi}{2}$ for type~II, which maps $\Acal$ to a dual configuration~$\Acal_\textrm{D}$
and likewise $\Fcal$ to $\Fcal_\textrm{D}$.

\section{Rational electromagnetic fields on Minkowski space}

\noindent
We have completely solved the vacuum Maxwell equations on the Lorentzian cylinder ${\cal I}\times S^3$ and, hence, on 
de Sitter space~dS$_4$. By conformal invariance, this solution carries over to any conformally equivalent spacetime.
In particular, the $\omega_4\equiv\cos\chi<\cos\tau$ half of the cylinder 
is mapped to the future half of Minkowski space~$\R^{1,3}_+\ni(t>0,x,y,z)$ via
\begin{equation} \label{transf2}
\cot\tau \= \frac{r^2 {-} t^2 {+} \ell^2}{2\,\ell\,t}\ ,\quad
\omega_1 \= \gamma\,\frac{x}{\ell}\ ,\quad
\omega_2 \= \gamma\,\frac{y}{\ell}\ ,\quad
\omega_3 \= \gamma\,\frac{z}{\ell}\ ,\quad
\omega_4 \= \gamma \frac{r^2 {-} t^2 {-} \ell^2}{2\,\ell^2}\ ,
\end{equation}
with the convenient abbreviations
\begin{equation}
r^2 \= x^2+y^2+z^2 \quad\und\quad
\gamma \= \frac{2\,\ell^2}{\sqrt{4\,\ell^2 t^2 + (r^2-t^2+\ell^2)^2}}
\=  \frac{2\,\ell^2}{\sqrt{4\,\ell^2 r^2 + (t^2-r^2+\ell^2)^2}}\ .
\end{equation}
The de Sitter metric~\eqref{dSmetric} in these coordinates becomes
\begin{equation}\label{metricMink}
\diff s^2 \=
\frac{\ell^2}{t^2}\,\bigl(-\mathrm{d}t^2 +\diff x^2 +\diff y^2 +\diff z^2\bigr) 
\quad\with (x,y,z)\equiv(x^1,x^2,x^3)\in\R^3 \und t\in\R_+ \ ,
\end{equation}
revealing its conformal equivalence to the Minkowski metric.
We may in fact cover the entire $\R^{1,3}$ by gluing on a second dS$_4$ copy at $t{=}\tau{=}0$, 
with $\tau\in(-\pi,0)$ but again restricted to $\cos\chi<\cos\tau$.
This doubles the Lorentzian cyclinder to $2{\cal I}\times S^3$ and extends the temporal range to $\tau\in(-\pi,\pi)$.
\begin{figure}[h!]
\centering
\includegraphics[width = 0.6\paperwidth]{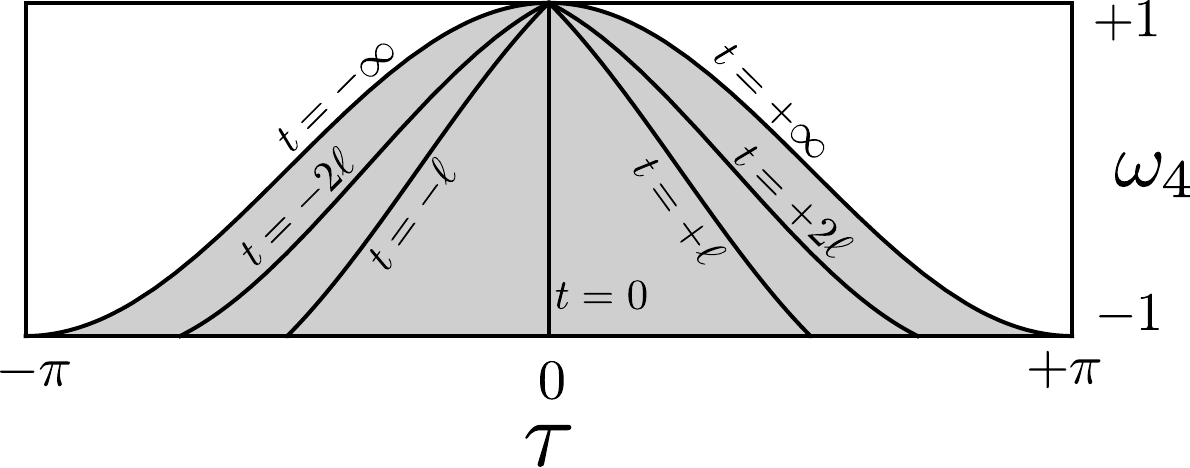}
\caption{An illustration of the map between a cylinder $2\mathcal{I}\times S^3$ and Minkowski space $R^{1,3}$.
The Minkowski coordinates cover the shaded area. The boundary of this area is given by the curve $\omega_4{=}\cos\chi{=}\cos\tau$. 
Each point is a two-sphere spanned by $\{\omega_1,\omega_2,\omega_3\}$, which is mapped to a sphere of constant $r$ and $t$. }
\end{figure}

The transformations \eqref{transf1} and~\eqref{transf2} give a map 
between Minkowski space and half of de Sitter space. 
Let us rewrite~\eqref{4vector} as
\begin{equation} \label{3vector}
\omega_i = \sin\chi\,\hat{x}^i  \und \omega_4 = \cos\chi \quad\with
(\hat{x}^i) = (\hat{x}^1,\hat{x}^2,\hat{x}^3) \= (\sin\theta\,\cos\phi\ ,\ \sin\theta\,\sin\phi\ ,\ \cos\theta)
\end{equation}
exposing the radial unit vector in~$\R^3$ parametrizing the $S^2$ slice at fixed~$\chi$.
On the other hand, $x^i=r\,\hat{x}^i$ in~\eqref{transf2} implies that we may identify the unit $S^2\ni(\theta,\phi)$ 
on both sides of the map after switching to spherical coordinates on Minkowski space,
\begin{equation}
(x^\mu)\=(t,x,y,z) \qquad\Rightarrow\qquad
(y^\rho)\=(t,r,\theta,\phi) \quad\for \mu,\rho=0,1,2,3\ .
\end{equation}
Thereby the map is reduced to one between $(\tau,\chi)$ and $(t,r)$,
\begin{equation} \label{transf2red}
\cos\tau \= \sfrac12\gamma\,(r^2-t^2+\ell^2)/\ell^2  \quad\und\quad
\cos\chi \= \sfrac12\gamma\,(r^2-t^2-\ell^2)/\ell^2 \ ,
\end{equation}
or
\begin{equation}
\sin\tau \= \gamma\,t/\ell \quad\und\quad \sin\chi \= \gamma\,r/\ell 
\qquad\Rightarrow\qquad \frac{\sin\chi}{\sin\tau} \= \frac{r}{t}\ .
\end{equation}
These relations are easily inverted to yield 
\begin{equation}
\gamma\= \cos\tau-\cos\chi\ >0
\end{equation}
and thus
\begin{equation} \label{transfinv}
\frac{t}{\ell} \= \frac{\sin\tau}{\cos\tau\,-\,\cos\chi} \quad\und\quad
\frac{r}{\ell} \= \frac{\sin\chi}{\cos\tau\,-\,\cos\chi} \qquad\for\quad
\chi>|\tau|\ .
\end{equation}
The triangular $(\tau,\chi)$ domain is nothing but the Penrose diagram of Minkowski space.
Special lines and points are
\begin{center}
\begin{tabular}{|c|ccc|ccc|c|}
\hline
& \quad$\tau{=}0$ & south pole & boundary & \quad --- \quad{} & \!\!\!north pole\!\!\! & $\quad\tau{=}{\pm}\pi\quad$ & $\chi{\pm}\tau{=}\pi$ \\[2pt]
$(\tau,\chi)$ & \quad$(0,\chi)$ & $(\tau,\pi)$ & $(\pm\chi,\chi)$ & $(0,\pi)$ & $(0,0)$ & $(\pm\pi,\pi)$ & $(\pm\pi{\mp}\chi,\chi)$ \\[4pt]
$(t,r)$ & \quad$(0,r)$ & $(t,0)$ & $(\pm\infty,\infty)$ & $(0,0)$ & $(t,\infty)$ & $(\pm\infty,r)$ & $(\pm r,r)$ \\[2pt]
& \quad$t{=}0$ & $r{=}0$ & $\mathscr{I}^\pm$ & \ \ origin \ {} & $i^0$ & $i^\pm$ & \ \  lightcone \ \ {} \\
\hline
\end{tabular}
\end{center}
where Minkowski spatial and temporal infinity $i^0$ and $i^\pm$ correspond to 
the corners of the Penrose diagram and are not included in the edges connecting them.
The behavior at the conformal boundary $\chi{=}|\tau|$ yields the properties at Minkowski null infinity~$\mathscr{I}^\pm$.
\begin{figure}[h!]
\centering
\includegraphics[width = 0.3\paperwidth]{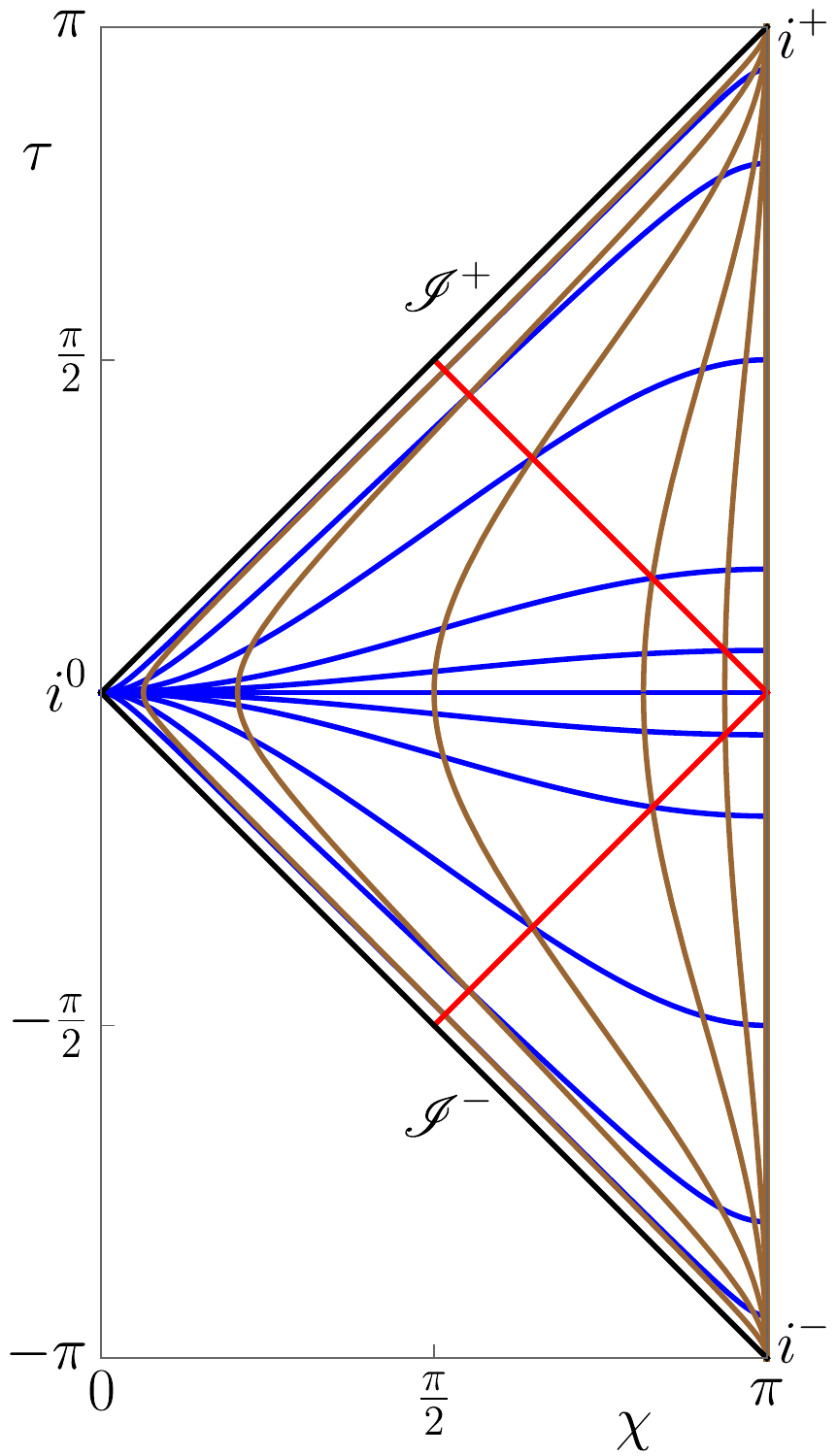}
\caption{
Penrose diagram of Minkowski space~$\R^{1,3}$. Each point hides a two-sphere $S^2\ni\{\theta,\phi\}$.
Blue curves indicate $t{=}\textrm{const}$ slices while brown curves depict the world volumes of $r{=}\textrm{const}$ spheres.
The lightcone of the Minkowski-space origin is drawn in red.
}
\end{figure}

We translate our Maxwell solutions from $2{\cal I}\times S^3$ to $\R^{1,3}$
simply by the coordinate change
\begin{equation}
\begin{aligned}
\tau=\tau(t,x,y,z) &\und \omega_{_A}=\omega_{_A}(t,x,y,z) \\[2pt]
\textrm{or}\qquad
\tau=\tau(t,r) &\und \chi=\chi(t,r)\ .
\end{aligned}
\end{equation}
In other words, abbreviating $x\equiv\{x^\mu\}$ and $y\equiv\{y^\rho\}$ and expanding
\begin{equation}
\begin{aligned}
\Acal &\= \Acal_a\bigl(\tau(x),\omega(x)\bigr)\,e^a(x) 
\= A_\mu(x)\,\diff x^\mu
\= A_\rho(y)\,\diff y^\rho \quad\und \\[4pt]
\diff\Acal &\= \pa_\tau\Acal_a\,e^\tau\we e^a + \bigl( R_b\Acal_c - \Acal_a\,\varepsilon^a_{\ bc}\bigr)\,e^b \we e^c 
\= \sfrac12 F_{\mu \nu}(x)\,\diff x^\mu \we \diff x^\nu 
\= \sfrac12 F_{\rho\lambda}(y)\,\diff y^\rho\we\diff y^\lambda
\end{aligned}
\end{equation}
we may read off $A_\mu$ (note that $A_t\neq0$~!) and 
$F_{\mu\nu}$ and thus the electric and magnetic fields
\begin{equation}
E_i \= F_{it} \quad\und\quad B_i \= \sfrac{1}{2}\varepsilon_{ijk}F_{jk}
\end{equation}
in Cartesian or in spherical coordinates.
To this end, we need to express our left-invariant one-forms~$e^a$ in terms of the Minkowski coordinates.
A straightforward but lengthy computation yields~\cite{lechtenfeld-zhilin}
\begin{equation}
\begin{aligned}
e^\tau &\= \sfrac{\gamma^2}{\ell^3}\bigl(
\sfrac12(t^2{+}r^2{+}\ell^2)\,\diff t - t\,x^k \diff x^k \bigr) \\[2pt]
&\= \sfrac{\gamma^2}{\ell^3}\bigl(
\sfrac12(t^2{+}r^2{+}\ell^2)\,\diff t - t\,r\,\diff r \bigr)
\qquad\und \\[4pt]
e^a &\= \sfrac{\gamma^2}{\ell^3}\bigl(
t\,x^a \diff t - \bigl[\sfrac12(t^2{-}r^2{+}\ell^2)\,\delta^{ak} +
x^a x^k + \ell\,\varepsilon^{ajk}x^j \bigr]\,\diff x^k \bigr) \\[2pt]
&\= \sfrac{\gamma^2}{\ell^3}\bigl(
\hat{x}^a \bigl[ r\,t\,\diff t -\sfrac12(t^2{+}r^2{+}\ell^2)\,\diff r\bigr]
- \sfrac12(t^2{-}r^2{+}\ell^2)\,r\,\diff\hat{x}^a -
\ell\,r^2\varepsilon^{ajk}\hat{x}^j\diff\hat{x}^k \bigr)\ .
\end{aligned}
\end{equation}
Alternatively, one may simply employ the Jacobian ($m\in\{\tau,\chi,\theta,\phi\}$)
\begin{equation} \label{Jacobian}
\bigl(J^m_{\ \ \mu}\bigr) \ :=\ \frac{\pa(\tau,\chi,\theta,\phi)}{\pa(t,r,\theta,\phi)}
\= \frac1\ell\biggl(\begin{matrix} p & -q \\[4pt] q & -p \end{matrix} \biggr) \oplus\mathds{1}_2
\quad\with\biggl\{ \begin{array}{l}
p\= \sfrac{\gamma^2}{\ell^2}\,(r^2{+}t^2{+}\ell^2)/2 \= 1{-}\cos\tau\cos\chi \\[4pt]
q\= \sfrac{\gamma^2}{\ell^2}\,t\,r \= \sin\tau\sin\chi \end{array}\biggr\}
\end{equation}
and
\begin{equation}
|\textrm{det}J| \= \frac{p^2{-}q^2}{\ell^2} \= \frac{\gamma^2}{\ell^2} 
\= \frac{\sin^2\!\tau}{t^2} \= \frac{\sin^2\!\chi}{r^2}
\quad\with \gamma^2=p^2{-}q^2
\end{equation}
to compute the spherical Minkowski components
\begin{equation}
A_t \= \Acal_\tau J^\tau_{\ t} + \Acal_\chi J^\chi_{\ t}  \ ,\qquad
A_r \= \Acal_\tau J^\tau_{\ r} + \Acal_\chi J^\chi_{\ r}  \ ,\qquad
A_\theta = \Acal_\theta \ ,\qquad A_\phi = \Acal_\phi\ ,
\end{equation}
and likewise any tensor component (we have gauged $\Acal_\tau{=}0$).
For later use, we also note here the transformation of the volume form
\begin{equation} \label{measure}
\diff^4x \= \diff t\,r^2\diff r\,\diff^2\Omega_2 \= r^2\,|\textrm{det}J|^{-1}\diff\tau\,\diff\chi\,\diff^2\Omega_2
\= \sin^2\!\chi\,|\textrm{det}J|^{-2}\diff\tau\,\diff\chi\,\diff^2\Omega_2
\=\frac{\ell^4}{\gamma^4}\,\diff\tau\,\diff^3\Omega_3\ .
\end{equation}

Furthermore, it comes in handy that $\Acal$ finally contains only even powers of~$\gamma$
and depends on~$\tau$ only through integral powers of
\begin{equation}
\exp (2\mathrm{i}\,\tau) \= \frac{[(\ell+\mathrm{i}t)^2+r^2]^2}{4\,\ell^2 t^2 + (r^2-t^2+\ell^2)^2}\ .
\end{equation}
Therefore, our Minkowski solutions have the remarkable property of being rational functions of~$(t,x,y,z)$.
More precisely, their electric and magnetic fields are of the form
\begin{equation}
\textrm{type I:} \quad \frac{P_{2(2j+1)}(x)}{Q_{2(2j+3)}(x)} \ ,\qquad
\textrm{type II:}\quad \frac{P_{2(2j-1)}(x)}{Q_{2(2j+1)}(x)} 
\end{equation}
where $P_r$ and $Q_r$ denote polynomials of degree~$r$.
Thus, as expected, their energy and action are finite.
Indeed, the fields fall off like $r^{-4}$ at spatial infinity for fixed time, but they decay
merely like $(t{\pm}r)^{-1}$ along the light-cone.
Hence, the asymptotic energy flow is concentrated on past and future null infinity~$\mathscr{I}^\pm$,
as it should be, but peaks on the light-cone of the spacetime origin.
Since on de Sitter space our basis solutions (\ref{type1}) and~(\ref{type2}) form a complete set, 
their Minkowski relatives are also complete in the space of finite-action configurations. 

For illustration, we display a type-I basis solution with $(j;m,n)=(1;0,0)$
obtained from 
\begin{equation}
\Acal_\pm \ \propto\ \sfrac{1}{\sqrt{2}}\,(\omega_1{\pm}\im\omega_2)(\omega_3{\pm}\im\omega_4)\,\cos 4\tau 
\qquad\und\qquad
\Acal_3\ \propto\  (\omega_1^2{+}\omega_2^2{-}\omega_3^2{-}\omega_4^2)\,\cos 4\tau\ .
\end{equation}
The resulting Riemann-Silberstein components are (up to overall scale)
{\small
\begin{equation}
\begin{aligned}
(E+\im B)_x &\= -\frac{2\im}{N}
\Bigl\{ 2y +3 \im t y -x z +2 t^2 y + 2\im t x z-8x^2 y-8y^3+4yz^2 \\
&\qquad+\ 4\im t^3 y -6t^2 xz -8\im t x^2 y-8\im t y^3+4\im t yz^2 +10x^3 z+10xy^2 z -2xz^3 \\
&\qquad+\ 2(\im t x z+x^2 y+y^3+y z^2)(-t^2{+}x^2{+}y^2{+}z^2)+(\im t y- x z) (-t^2{+}x^2{+}y^2{+}z^2)^2\Bigr\}
\ ,\quad{}
\end{aligned}
\end{equation}
\begin{equation}
\begin{aligned}
(E+\im B)_y &\= \frac{2\im}{N}
\Bigl\{ 2x +3 \im t x + y z+2 t^2 x-2\im t y z-8x^3-8xy^2 +4 xz^2 \\
&\qquad+\ 4 \im t^3 x +6 t^2y z -8\im t x^3-8\im txy^2 +4 \im t x z^2-10 x^2 yz-10y^3 z +2y z^3 \\
&\qquad+\ 2 (-\im t y z+x^3+x y^2+x z^2)(-t^2{}+x^2{+}y^2{+}z^2) +(\im t x+ y z)(-t^2{+}x^2{+}y^2{+}z^2)^2\Bigr\}
\ ,
\end{aligned}
\end{equation}
\begin{equation}
\begin{aligned}
(E+\im B)_z &\= \frac{\im}{N}
\Bigl\{1+2\im t+t^2-11 x^2- 11 y^2 +3 z^2 +4 \im t^3-16\im t x^2-16\im t y^2+4\im t z^2 \\
&\qquad-\ t^4-2 t^2 x^2-2t^2 y^2-2t^2 z^2+11 x^4+22x^2y^2-10 x^2 z^2+11y^4-10 y^2 z^2+3z^4 \\
&\qquad+\ 2 \im t (t^2{-}3 x^2{-}3 y^2{-}z^2) (t^2{-}x^2{-}y^2{-}z^2)-(t^2{+}x^2{+}y^2{-}z^2)(-t^2{+}x^2{+}y^2{+}z^2)^2\Bigr\}
\ ,\quad{}
\end{aligned}
\end{equation}
}
\begin{equation}
\textrm{with} \qquad
N=\bigl((t{-}\im)^2-x^2-y^2-z^2\bigr)^5\ . 
\qquad\qquad\qquad\qquad\qquad\qquad\qquad\qquad\qquad\qquad\qquad\qquad{}
\end{equation}

\section{Symmetry analysis}

\noindent
The main advantage of constructing Minkowski-space electromagnetic field configurations 
via the detour over de Sitter space is the enhanced manifest symmetry of our construction. 
The isometry group SO(1,4) of dS$_4$ is generated by 
($A,B=1,2,3,4$ and $a,b,c=1,2,3$, abbreviate $\frac{\partial}{\partial q_{_B}}\equiv\partial_{_B}$)
\begin{equation}
\{\Mcal_{_{AB}}{\equiv}\,{-}q_{[{\scriptscriptstyle A}}\partial_{{\scriptscriptstyle B}]}\,,\ 
\Mcal_{0{\scriptscriptstyle B}}{\equiv}\,q_{({\scriptstyle 0}}\partial_{{\scriptscriptstyle B})} \} \= 
\{ \Mcal_{ab}{=}\varepsilon_{abc}\Dcal_c\,,\ \Mcal_{4a}{=}\Pcal_a\,,\ \Mcal_{04}{=}\Pcal_0\,,\ \Mcal_{0b}{=}\Kcal_b\}\ ,
\end{equation}
which can be contracted (with $\ell{\to}\infty$) to the isometry group ISO(1,3) of $\R^{1,3}$ (the Poincar\'e group)
generated by ($\mu,\nu=0,1,2,3$ and $i,j,k=1,2,3$)
\begin{equation}
\{M_{\mu\nu}\,,\ P_\mu \} \=
\{ M_{ij}{=}\varepsilon_{ijk}D_k\,,\ P_i\,,\ P_0\,,\ M_{0j}{=}K_j\}\ ,
\end{equation}
where the two sets are ordered likewise, 
and we employ (as aleady earlier) calligraphic symbols for de Sitter quantities and straight symbols for Minkowskian ones.
Here, $D$ denotes spatial rotations, $P$ are translations, and $K$ stand for boosts in Minkowski space.

Since the two spaces are conformally equivalent already at $\ell{<}\infty$ via~\eqref{transf2},
the corresponding generators should be related. Indeed, the common SO(3) subgroup in
\begin{equation}
SO(1,4) \supset SO(4) \supset SO(3) \quad\und\quad ISO(1,3) \supset SO(1,3) \supset SO(3)
\end{equation}
is identified, $\Dcal_i{=}D_i{=}{-}2\varepsilon_{ij}^{\ k}x^j\partial_k$. 
However, any other generator becomes nonlinearly realized when 
mapped to the other space via \eqref{transf2red} or~\eqref{transfinv}.
For example, the would-be translation $\Pcal_3$ defined in~\eqref{DPdef} reads
\begin{equation}
\begin{aligned}
\Pcal_3 = L_3-R_3 &\= -2\,\cos\theta\,\pa_\chi + 2\,\cot\chi\sin\theta\,\pa_\theta \\[2pt]
&\= \sfrac1{\ell}\,\cos\theta\,\bigl( 2\,r\,t\,\pa_t + (t^2{+}r^2{+}\ell^2)\,\pa_r \bigr) 
- \sfrac1{\ell\,r} (t^2{-}r^2{+}\ell^2)\,\sin\theta\,\pa_\theta \\[2pt]
&\ \to\ 2\ell\,\bigl( \cos\theta\,\pa_r - \sfrac1r\,\sin\theta\,\pa_\theta \bigr)
\= 2\ell\,\pa_z \= \ell\,P_z \quad\for\ell\to\infty
\end{aligned}
\end{equation}
as it should be. Similarly, ${\cal P}_0\to \ell\,P_0$ and ${\cal K}_b\to K_j$ for $\ell{\to}\infty$ 
when expanded around $(t,r)=(\ell,0)$ corresponding to the $S^3$ south pole at $q_0{=}0$.
Nevertheless, the de Sitter construction enjoys an SO(4) covariance (generated by $\Dcal_a$ and $\Pcal_a$)
which extends the obvious SO(3) covariance in Minkowski space.
It allows us to connect all solutions of a given type (I or II) with a fixed value of the spin~$j$
by the action of SO(4) ladder operators $L_\pm$ and $R_\pm$ or $\Dcal_\pm$ and $\Pcal_a$, 
which is non-obvious on the Minkowski side. 
On the other hand, Minkowski boosts and translations have no simple realization on de Sitter space.

Actually, Maxwell theory on either space is also invariant under conformal transformations.
These may be generated by the isometry group together with a conformal inversion.
On the Minkowski side, the latter is
\begin{equation}
J : \quad x^\mu \ \mapsto\ \frac{x^\mu}{x\cdot x} \quad\with x\cdot x=r^2-t^2\ .
\end{equation}
We have to distinguish two cases:
\begin{equation}
\begin{aligned}
\textrm{spacelike:}\quad & t^2<r^2 \quad\Rightarrow\quad
J_> : \quad\bigl(t,r,\theta,\phi\bigr) \ \mapsto\ \bigl(\sfrac{t}{r^2-t^2}, \sfrac{r}{r^2-t^2},\theta,\phi\bigr)\ ,\\[2pt]
\textrm{timelike:} \quad & t^2>r^2 \quad\Rightarrow\quad
J_< : \quad\bigl(t,r,\theta,\phi\bigr) \ \mapsto\ \bigl(\sfrac{-t}{t^2-r^2}, \sfrac{r}{t^2-r^2},\pi{-}\theta,\phi{+}\pi\bigr)\ .
\end{aligned}
\end{equation}
On the de Sitter side, this is either (spacelike) a reflection on the $S^3$ equator~$\chi{=}\frac{\pi}{2}$
or (timelike) a $\pi$-shift in cylinder time~$\tau$ plus an $S^2$ antipodal flip,
\begin{equation}
\begin{aligned}
\textrm{spacelike:}\quad & |\tau|{+}\chi<\pi \quad\Rightarrow\quad
\Jcal_> :  \quad \bigl(\tau,\chi,\theta,\phi\bigr)\ \mapsto\ \bigl(\tau,\pi{-}\chi,\theta,\phi\bigr)\ , \\[2pt]
\textrm{timelike:} \quad & |\tau|{+}\chi>\pi \quad\Rightarrow\quad
\Jcal_< :  \quad \bigl(\tau,\chi,\theta,\phi\bigr)\ \mapsto\ \bigl(\tau{\pm}\pi,\chi,\pi{-}\theta,\phi{+}\pi\bigr)\ .
\end{aligned}
\end{equation}
In the spacelike case, merely the sign of $\omega_4{\equiv}\cos\chi$ gets flipped, 
which amounts to a parity flip $L\leftrightarrow R$. 
In the timelike case, both $\cos\tau$ and $\sin\tau$ change sign, 
which combines a time reversal with a reflection at $\tau{=}\frac{\pi}{2}$ or $\tau{=}{-}\frac{\pi}{2}$.
Note that it is different from the $S^3$ antipodal map, which
is not a reflection but a proper rotation, $\omega_{_A}\mapsto-\omega_{_A}$ or 
$(\chi,\theta,\phi)\mapsto(\pi{-}\chi,\pi{-}\theta,\phi{+}\pi)$.
The lightcone is singular under the inversion; it is mapped to the conformal boundary
$r{=}{\pm}t{=}\infty$ or $\chi{=}{\pm}\tau$.
We infer that the conformal inversion allows us to relate type-I and type-II solutions of the same spin.
It is easily checked that the spatial fall-off behavior of our rational solutions is not modified by the inversion.

Finally, one may consider dilatations in Minkowski space,
\begin{equation}
x^\mu \ \mapsto\ \lambda\,x^\mu \quad\for \lambda\in\R_+\ .
\end{equation}
However, this amounts to a trivial rescaling also achieved by changing the de Sitter radius,
$\ell\mapsto\lambda\,\ell$, as the scale~$\ell$ was removed on the Lorentzian cylinder.

\section{Energy and helicity}

\noindent
The Maxwell system features two conserved quantities, the field energy~$E$ and its helicity~$h$. 
Both are given by spatial integrals, but the choice of time slice is inconsequential due to the conservation.
It is most convenient to pick the $t=\tau=0$ slice. The energy is then given by~\cite{lechtenfeld-zhilin}
\begin{equation}\label{energy}
E = E_{\textrm{el}}+E_{\textrm{mag}} \= \sfrac12 \int_{\R^3} \!\diff^3\! x \ \bigl(\vec{E}^2 + \vec{B}^2\bigr) \= 
\sfrac{1}{2\ell}\int_{S^3} \diff^3\Omega_3 \  (1-\cos\chi)\,\bigl({\cal E}_a{\cal E}_a + {\cal B}_a{\cal B}_a\bigr)\ .
\end{equation}
(The orientation of the $S^3$ volume measure~$\diff^3\Omega_3$ is chosen to provide a positive result.)
By recalling \eqref{Asep} and \eqref{cur-field} and suppressing the index~$j$ for a fixed spin value and solution type one has
\begin{equation} \label{EBj}
\Acal_a \= X_a(\omega)\,\ep^{\Omega\,\im\tau} + \bar{X}_a(\omega)\,\ep^{-\Omega\,\im\tau}
\qquad\Rightarrow\qquad \biggl\{ \begin{array}{l}
\Ecal_a \= -\im\,\Omega\,X_a\,\ep^{\Omega\,\im\tau} + \im\,\Omega\,\bar{X}_a\,\ep^{-\Omega\,\im\tau} \\[2pt]
\Bcal_a \= \,\mp\,\Omega\,X_a\,\ep^{\Omega\,\im\tau}\,\mp\,\Omega\,\bar{X}_a\,\ep^{-\Omega\,\im\tau} \end{array} \biggr\}
\end{equation}
with $\bar{X}_a$ denoting the complex conjugate of $X_a$. 
Thus we obtain a time-independent ``sphere-frame'' energy density
\begin{equation}
\sfrac12 \bigl( \Ecal_a \Ecal_a + \Bcal_a \Bcal_a \bigr) \= 2\,\Omega^2\,X_a\bar{X}_a(\omega)
\end{equation}
with $\Omega=\Omega(j)$ for either solution type and fixed spin~$j$.
The total energy can then be computed from \eqref{energy} by using the harmonic expansion of $X_a^j(\omega)$ 
as obtained previously through \eqref{XZ}, \eqref{basisZ}, \eqref{type1} and~\eqref{type2} 
while making use of the orthogonality of the left-right harmonics $Y_{j;m,n}$
\begin{equation}
\int_{S^3} \!\diff^3\Omega_3\ Y_{j;m,n}\,\overline{Y}_{j';m',n'} \= \delta_{jj'}\delta_{mm'}\delta_{nn'}\ ,
\end{equation}
to obtain
\begin{equation}
E^{(j)} \= \sfrac1{\ell}\,(2j{+}1)\,\Omega^3\sum\limits_{m,n}|\lambda_{mn}|^2\ .
\end{equation}

The expression for the helicity is metric-free and can thus be evaluated over any spatial slice. 
Choosing again $t=\tau=0$,
\begin{equation} \label{helicity}
h = h_{\textrm{mag}}+h_{\textrm{el}} \= \sfrac12 \int_{\R^3} \ \bigl( A\wedge F + A_D \wedge F_D \bigr)
\= -\sfrac12 \int_{S^3} \diff^3\Omega_3 \  (1-\cos\chi)\,\bigl(\Acal_a\Bcal_a + \Acal^D_a\Ecal_a\bigr)\ .
\end{equation}
Once again, taking type~I (upper sign) or type~II (lower sign) and fixing the spin~$j$ we obtain
\begin{equation}
\Acal^D_a \= \pm\im\,X_a(\omega)\,\ep^{\Omega\,\im\tau} \mp\im\,\bar{X}_a(\omega)\,\ep^{-\Omega\,\im\tau}\ ,
\end{equation}
which yields a constant ``sphere-frame'' helicity density
\begin{equation}
-\sfrac12 \bigl( \Acal_a\Bcal_a + \Acal^D_a\Ecal_a \bigr) \= \pm 2\,\Omega\,X_a\bar{X}_a(\omega)\ .
\end{equation}
As a result, even before performing the $S^3$ integration, we find a linear helicity-energy relation 
\begin{equation}
\Omega\, h \= \pm \ell\, E \qquad\textrm{for fixed spin and type}\ .
\end{equation}

Since the helicity measure an average of the linking numbers of any two electric or magnetic field lines~\cite{moffatt, berger},
the latter must be related to the value~$j$ of the spin. The individual linking number of two field lines, however, appears neither to be independent of the lines chosen nor constant in time, as our observations indicate.
An exception are the Ra\~nada--Hopf knots $(j{=}m{=}0,n{=}{\pm}1)$, 
which display a conserved linking number of unity between any pair of electric
or magnetic field lines.

\section{Null fields}

\noindent
An interesting subset of vacuum electromagnetic fields are those with vanishing Lorentz invariants,
\begin{equation}
\vec{E}^2-\vec{B}^2 =0 \quad\und\quad \vec{E}\cdot\vec{B} =0 \qquad\Longleftrightarrow\qquad
\bigl(\vec{E}\pm\im\vec{B}\bigr)^2 =\ 0\ .
\end{equation}
As a scalar equation it must equally hold on the de Sitter side, and so
we can try to characterize such configurations with our SO(4) basis above. For a given type and spin, 
the expressions in~\eqref{EBj} immediately give the Riemann-Silberstein vector on the $S^3$~cylinder,
\begin{equation}
\Ecal_a \pm\im\,\Bcal_a \= -2\im\,\Omega\,X_a(\omega)\,\ep^{\Omega\,\im\tau}\ ,
\end{equation}
where the upper (lower) sign pertains to type~I (II).
Note that the negative-frequency part of this field has cancelled.
The vanishing of $(\Ecal_a{\pm}\im\Bcal_a)(\Ecal_a{\pm}\im\Bcal_a)$ is then equivalent to a condition on the angular functions,
\begin{equation} \label{nullcondition}
0 \= X_1(\omega)^2 + X_2(\omega)^2 + X_3(\omega)^2 \= 2\,Z_+(\omega) Z_-(\omega) + Z_3(\omega)^2\ .
\end{equation}
When expanding the angular functions~$Z^j_{*\ \textrm{I}}$ or $Z^j_{*\ \textrm{II}}$ into basis solutions according to~\eqref{basisZ},
one arrives at a system of homogeneous quadratic equations for the free coefficients~$\lambda_{j;m,n}^{\textrm{I/II}}$.

Let us analyze the situation for type~I and spin~$j$. 
The functions~$Z^j_*(\omega)$ transform under a $(j,j)$ representation of $su(2)_L\oplus su(2)_R$. 
The null condition~\eqref{nullcondition} then yields a representation content of $(0,0)\oplus(1,1)\oplus\ldots\oplus(2j,2j)$
and may thus be expanded into the corresponding harmonics. The independent vanishing of all coefficients produces
$\frac16(4j{+}1)(4j{+}2)(4j{+}3)$ equations for the $(2j{+}1)(2j{+}3)$ parameters~$\lambda_{j;m,n}$ (note the ranges of $m$ and~$n$ for type~I).
Clearly, this system is vastly overdetermined. However, it turns out that only $4j^2{+}6j{+}1$ equations are independent,
still leaving $2j{+}2$ free complex parameters for the solution space. The independent equations can be organized as (suppressing~$j$)
\begin{equation}
\begin{aligned}
\lambda_{m,n}^2 &\ \sim\ \lambda_{m,n-1}\,\lambda_{m,n+1} \qquad\ \for m,n=-j\,\ldots,j \ ,\\
\lambda_{m,j+1}\,\lambda_{m+1,-j-1} &\= \lambda_{m+1,j+1}\,\lambda_{m,-j-1} \quad\for m=-j,\ldots,j{-}1\ .
\end{aligned}
\end{equation}
We have checked for $j{\le}5$ that the upper equations are solved by~\footnote{
We thank Colin Becker for the verification. These are the generic solutions.  There exist also special solutions given by 
\eqref{extweights} and $\lambda_{m,n}=0$ for $|n|\neq j{+}1$, for arbitrarily selected choices of~$m\in\{-j,\ldots,j\}$. }
\begin{equation}
\lambda_{m,n}^{2j+2} \= {\textstyle\sqrt{\binom{2j+2}{j+1-n}}}\ \lambda_{m,-j-1}^{j+1-n}\,\lambda_{m,j+1}^{j+1+n}
\quad\for m=-j,\ldots,j \und n=-j{-}1,\ldots,j{+}1\ ,
\end{equation}
while the lower ones imply that the highest weights $n{=}j{+}1$ and the lowest weights $n{=}{-}j{-}1$
are proportional to one another (independent of~$m$), 
\begin{equation} \label{extweights}
\lambda_{m,-j-1} \= w\,\lambda_{m,j+1} \quad\for w\in\C^*\ .
\end{equation}
Therefore, the full (generic) solution reads
\begin{equation}
\lambda_{m,n} \= {\textstyle\sqrt{\binom{2j+2}{j+1-n}}}\ w^{\frac{j+1-n}{2j+2}}\ \ep^{2\pi\im k_m\frac{j+1-n}{2j+2}}\ z_m 
\qquad\with z_m\in\C \und k_m\in\{0,1,\ldots,2j{+}1\}\ ,
\end{equation}
containing $2j{+}2$ complex parameters $z_m$ and~$q$ as well as $2j$ discrete choices~$\{k_m\}$ 
(one of them can be absorbed into~$z_m$). This completely specifies the type-I null fields for a given spin.
Type-II null fields are easily obtained by applying electromagnetic duality to type-I null fields.

In the simplest case of $j{=}0$, the single equation $\lambda_{0,0}^2=2\lambda_{0,-1}\lambda_{0,1}$ describes 
a generic rank-3 quadric in $\C P^2$, or a cone over a sphere $\C P^1$ inside the parameter space~$\C^3$. For higher spin,
the moduli space of type-I null fields remains a complete-intersection projective variety of complex dimension~$2j{+}1$.~\footnote{
O.L.~is grateful to Harald Skarke for clarifications on this issue.}

We conclude the Section with a display of energy densities for a type-I $j{=}\sfrac12$ and $j{=}1$ null field at $t{=}0$
together with typical field lines of these basis solutions.
For $t{\neq}0$ the pictures get smoothly distorted.
\begin{figure}[h!]
\centering
\includegraphics[width = 0.35\paperwidth]{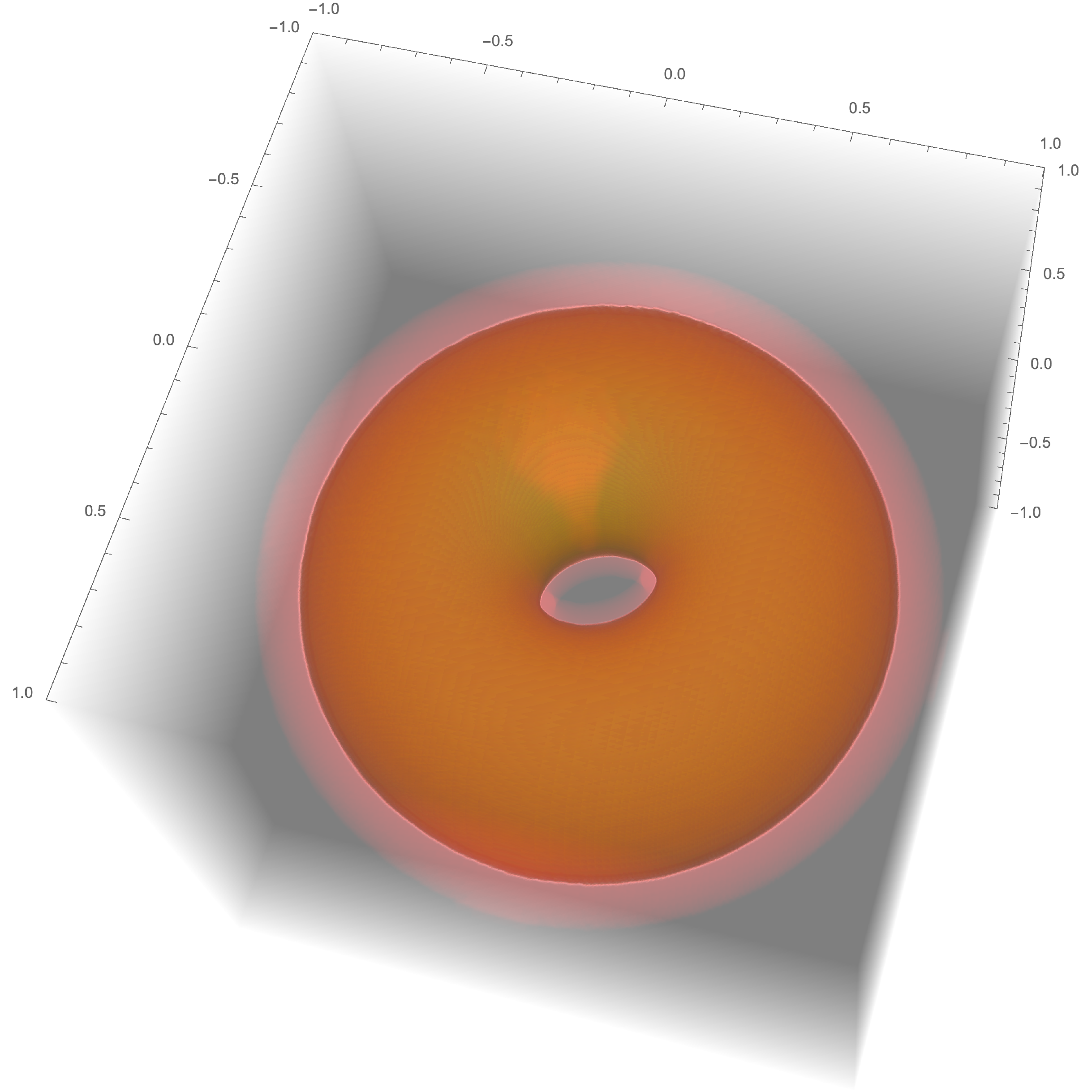}
\qquad
\includegraphics[width = 0.35\paperwidth]{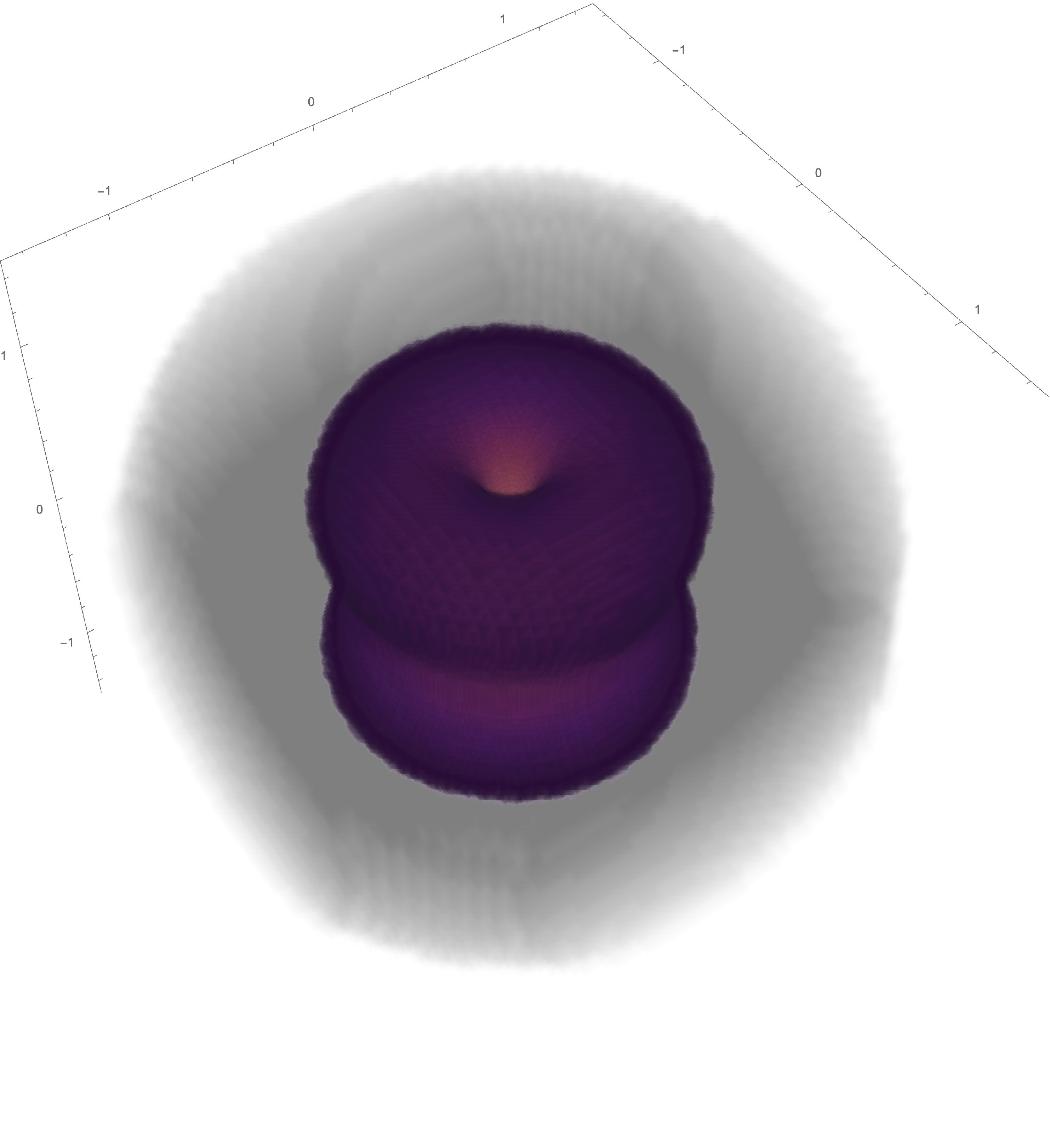}
\caption{
Energy density at $t{=}0$ for $(j;m,n)=(\sfrac12;\sfrac12,\sfrac32)$ (left) and $(1;0,2)$ (right) basis solutions.
}
\end{figure}
\begin{figure}[h!]
\centering
\includegraphics[width = 0.35\paperwidth]{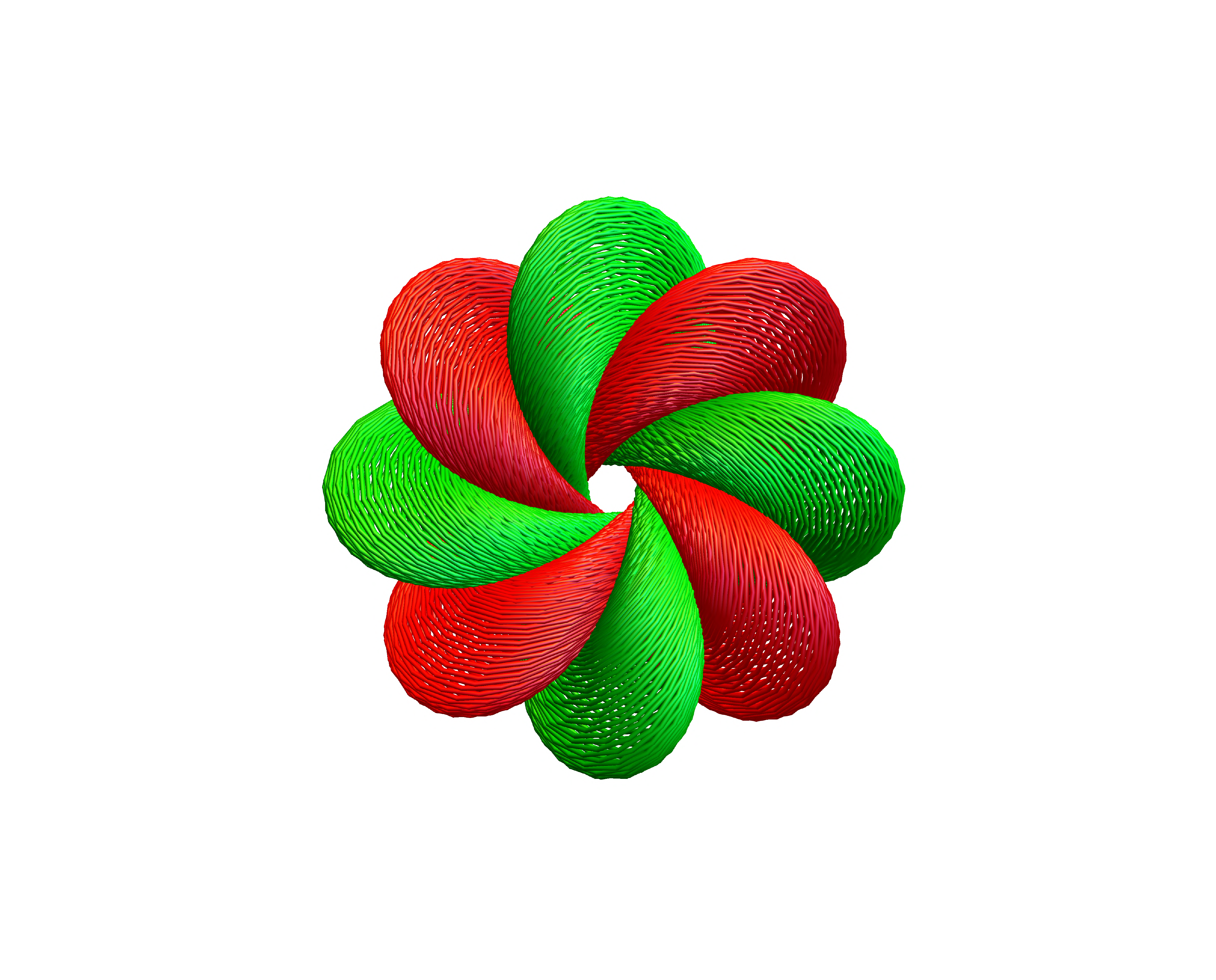}
\qquad
\includegraphics[width = 0.35\paperwidth]{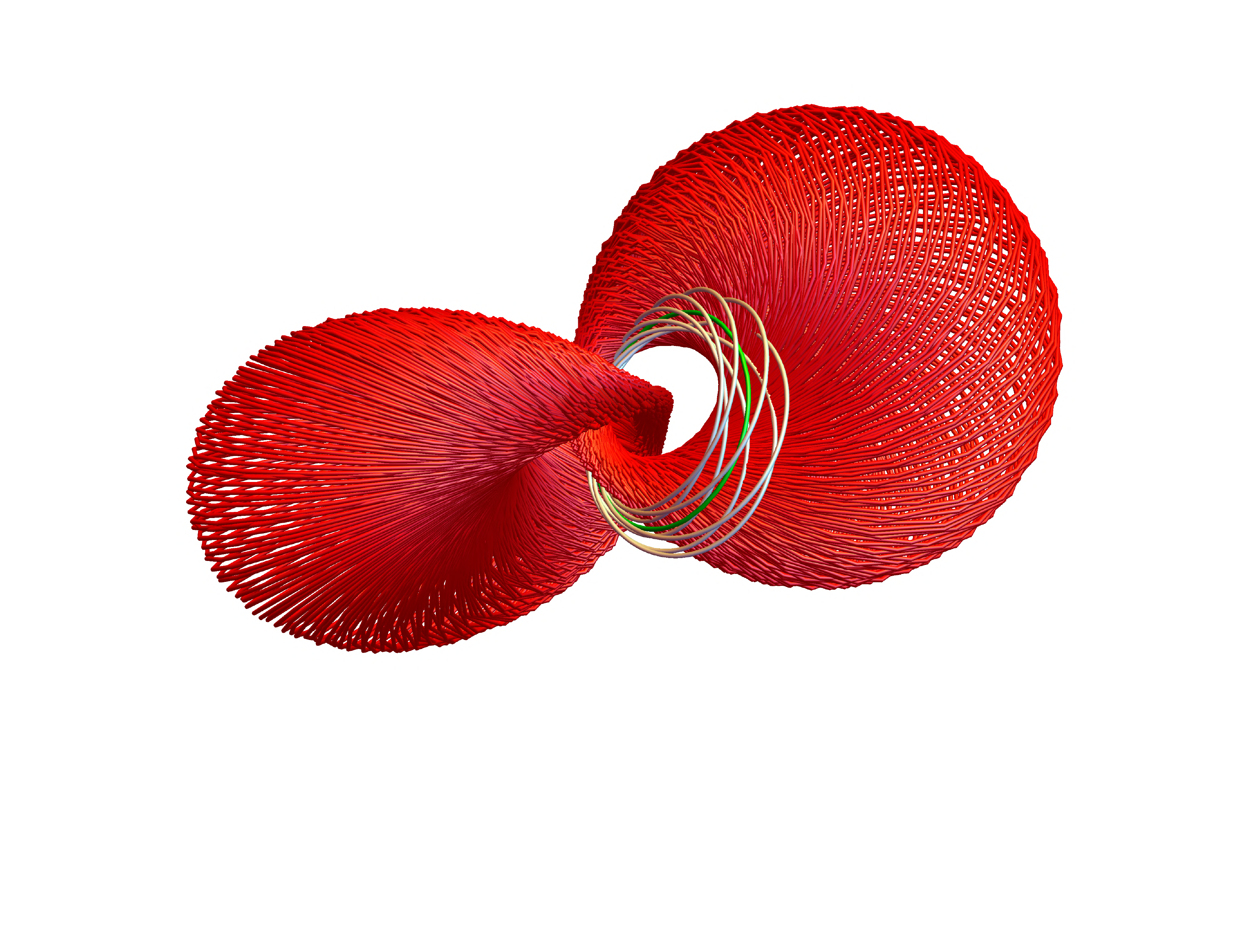}
\caption{
Sample electric (red) and magnetic (green) field lines at $t{=}0$ of the configuration in the corresponding figure above.
Left: a pair of electric and a pair of magnetic field lines. Right: a pair of electric field lines, a magnetic field line
of self-linking one and a magnetic field line of self-linking seven.
}   
\end{figure}

\section{Electromagnetic flux at infinity}

\noindent
We have seen that electromagnetic energy is radiated away along the light-cones. 
Let us try to quantify its amount over future null infinity~$\mathscr{I}^+$.
The energy flux at time~$t_0$ passing through a two-sphere of radius~$r_0$ centered at the spatial origin is given by
\begin{equation}
\Phi(t_0,r_0) \= \int_{S^2(r_0)}\!\!\diff^2\vec{\sigma}\cdot \bigl(\vec{E}\times\vec{B}\bigr)(t_0,r_0,\theta,\phi) 
\= \int_{S^2}r_0^2\,\diff^2\Omega_2\ T_{t\,r}^{\mathrm{(M)}}(t_0,r_0,\theta,\phi)\ ,
\end{equation}
where $\diff^2\Omega_2=\sin\theta\,\diff\theta\,\diff\phi$, 
and $T_{t\,r}^{\mathrm{(M)}}$ is the $(t,r)$ component of the Minkowski-space stress-energy tensor
\begin{equation}\label{stress-energy}
T_{\mu\,\nu}^{\mathrm{(M)}} \= F_{\mu\rho}F_{\nu\lambda}\,g^{\rho\lambda} - \sfrac14 g_{\mu\nu}F^2 
\quad\with (g_{\mu\nu}) = \textrm{diag}(-1,1,r^2,r^2\sin^2\!\theta) \for \mu,\nu,\ldots\in\{t,r,\theta,\phi\}\ .
\end{equation}
We carry out this computation in the $S^3$-cylinder frame by using the conformal relations
\begin{equation}
\ell^2\,T_{\mu\,\nu}^{\mathrm{(dS)}} \= t^2\,T_{\mu\,\nu}^{\mathrm{(M)}} \= \sin^2\!\tau\,T_{\mu\,\nu}^{\mathrm{(cyl)}} 
\= \sin^2\!\tau\,T_{m\,n}^{\mathrm{(cyl)}}\, J^m_{\ \ \mu}\,J^n_{\ \ \nu} 
\quad\for m,n\in\{\tau,\chi,\theta,\phi\}
\end{equation}
with the Jacobian~\eqref{Jacobian} and the fact that \ $r\sin\tau=t\sin\chi$ \ so that 
\begin{equation}
\Phi(\tau_0,\chi_0) \= \int_{S^2}\sin^2\!\chi\,\diff^2\Omega_2\ T_{t\,r}^{\mathrm{(cyl)}}(\tau_0,\chi_0,\theta,\phi)
\=  \int_{S^2}\sin^2\!\chi\,\diff^2\Omega_2\ T_{m\,n}^{\mathrm{(cyl)}}\, J^m_{\ \ t}\,J^n_{\ \ r} \ .
\end{equation}
A straightforward computation using $(g_{mn})=\textrm{diag}(-1,1,\sin^2\!\chi,\sin^2\!\chi\sin^2\!\theta)$ then yields
\begin{equation}
\begin{aligned}
\Phi(\tau_0,\chi_0) &\= \frac{p\,q}{\ell^2}\int \diff^2\Omega_2\,\Bigl((\Fcal_{\tau\theta})^2 + (\Fcal_{\chi\theta})^2 
+ \sfrac{1}{\sin^2\!\theta}\bigl[(\Fcal_{\tau\phi})^2 + (\Fcal_{\chi\phi})^2\bigr]\Bigr) \\
&\quad +\ \frac{p^2{+}q^2}{\ell^2}\int \diff^2\Omega_2\,\Bigl( \Fcal_{\tau\theta}\,\Fcal_{\chi\theta} 
+ \sfrac{1}{\sin^2\!\theta}\Fcal_{\tau\phi}\,\Fcal_{\chi\phi}\Bigr)\ .
\end{aligned}
\end{equation}
The sphere-frame components $\Fcal_{mn}$ can be computed by expanding
$e^a=e^a_{\ m}\,\diff\xi^m$ in
\begin{equation}
\Fcal \=  \Ecal_a\,e^a \we e^\tau + \sfrac12\,\Bcal_a\,\varepsilon^a_{\ bc}\,e^b\we e^c \= \Fcal_{mn}\,\diff\xi^m\we\diff\xi^n
\quad\with\xi^n\in\{\tau,\chi,\theta,\phi\}\ .
\end{equation}
The expression for the flux in sphere-frame fields then becomes
\begin{equation}\label{flux}
\begin{aligned}
\ell^2\,\Phi &\= p\,q\,\sin^2\!\chi \int_{S^2}\diff^2\Omega_2\ \Bigl[
(\sin\phi\,\Ecal_1-\cos\phi\,\Ecal_2)^2 + (\cos\theta\cos\phi\,\Ecal_1+\cos\theta\sin\phi\,\Ecal_2-\sin\theta\,\Ecal_3)^2 \\
&\qquad\qquad\qquad\qquad\quad +\ (\sin\phi\,\Bcal_1-\cos\phi\,\Bcal_2)^2 + (\cos\theta\cos\phi\,\Bcal_1+\cos\theta\sin\phi\,\Bcal_2-\sin\theta\,\Bcal_3)^2 \Bigr] \\
&\ +\ (p^2{+}q^2)\,\sin^2\!\chi \int_{S^2}\diff^2\Omega_2\ \Bigl[
(\sin\phi\,\Bcal_1-\cos\phi\,\Bcal_2) (\cos\theta\cos\phi\,\Ecal_1+\cos\theta\sin\phi\,\Ecal_2-\sin\theta\,\Ecal_3) \\
&\qquad\qquad\qquad\qquad\qquad\quad -\ (\sin\phi\,\Ecal_1-\cos\phi\,\Ecal_2) (\cos\theta\cos\phi\,\Bcal_1+\cos\theta\sin\phi\,\Bcal_2-\sin\theta\,\Bcal_3) \Bigr]\ .
\end{aligned}
\end{equation}

The total energy flux across future null infinity is obtained by evaluating this expression on $\mathscr{I}^+$ and integrating over it. Introducing cylinder light-cone coordinates
\begin{equation}
u=\tau{+}\chi \und v=\tau{-}\chi \qquad\textrm{so that}\qquad
t{+}r=-\ell\,\cot\sfrac{v}{2} \und t{-}r=-\ell\,\cot\sfrac{u}{2}
\end{equation}
we characterize $\mathscr{I}^+$ as
\begin{equation}
\biggl\{ \begin{array}{l} t{+}r\to\infty \\[4pt]  t{-}r \in\R \end{array} \biggr\} \quad\Leftrightarrow\quad
\biggl\{ \begin{array}{l} u\in(0,2\pi) \\[4pt]  v=0 \end{array} \biggr\} \qquad\Rightarrow\qquad
p=q=\sin^2\!\chi \und \gamma=0\ .
\end{equation}
Further noticing that
\begin{equation}
\diff(t{-}r) \= \frac{\ell\ \diff u}{p{+}q} \= \frac{\ell\ \diff u}{1-\cos u} \quad\und\quad
\sin^2\!\chi \= \sin^2\!\sfrac{u-v}{2} \= \sfrac12\bigl(1-\cos(u{-}v)\bigr)\ ,
\end{equation}
we may express this total flux as
\begin{equation}
\Phi_+ \= \int_{-\infty}^\infty \diff(t{-}r)\ \Phi\big|_{\mathscr{I}^+}
\= \int_0^{2\pi} \frac{\ell\ \diff u}{1-\cos u}\ \Phi(\sfrac{u}{2},\sfrac{u}{2})
\end{equation}
to obtain
\begin{equation} \label{totalflux}
\begin{aligned}
\Phi_+ &\=\frac{1}{8\ell} \int\!\diff u\,(1{-}\cos u)^2 \int\!\diff^2\Omega_2\ \Bigl[
\bigl\{ \cos\theta\cos\phi\,\Ecal_1+\cos\theta\sin\phi\,\Ecal_2-\sin\theta\,\Ecal_3+\sin\phi\,\Bcal_1-\cos\phi\,\Bcal_2 \bigr\}^2 \\
&\qquad\qquad\qquad\qquad\qquad\qquad\ \
+\ \bigl\{ \cos\theta\cos\phi\,\Bcal_1+\cos\theta\sin\phi\,\Bcal_2-\sin\theta\,\Bcal_3-\sin\phi\,\Ecal_1+\cos\phi\,\Ecal_2 \bigr\}^2 \Bigr]
\end{aligned}
\end{equation}
The square bracket expression above can be further simplified for a fixed spin and type 
by employing~\eqref{EBj} along with \eqref{XZ}, \eqref{basisZ}, \eqref{type1} and~\eqref{type2} to get
\begin{equation}
\Phi_+^{(j)} \= \frac{\Omega^2}{4\ell}\int\!\diff u\,(1{-}\cos u)^2 \int\!\diff^2\Omega_2\,
 \big| \pm Z_+^j e^{-\im\phi}(1\pm \cos\theta)\mp Z_-^j e^{\im\phi}(1\mp \cos\theta)-\sqrt{2}Z_3^j\sin\theta\, \big|^2\ ,
\end{equation}
where the upper (lower) sign corresponds to a type-I (type-II) solution. 
In the special case of $j=0$ $(\Omega{=}2)$ the contribution to the two-sphere integral only comes 
from the part which is independent of $(\theta,\phi)$, i.e.~$\frac{4}{3}\left(|Z^0_+|^2+|Z^0_-|^2+|Z^0_3|^2\right)$, 
so that the integration can easily be performed by passing to the adjoint harmonics 
$\tilde{Y}_{j;l,M}$ \eqref{Y-new} and using \eqref{split-Y}  to get
\begin{equation}
\Phi_+^{(0)} \= \frac{16}{3\,\ell}\int\limits_0^{2\pi}\!\diff u\ \sin^4\!\sfrac{u}{2}\,
\bigl| R_{0,0}(\sfrac{u}{2})\bigr|^2\sum\limits_{n=-1}^{1}|\lambda_{0,n}|^2 
\= \frac{8}{\ell}\sum\limits_{n=-1}^{1}|\lambda_{0,n}|^2 \= E^{(0)}\ .
\end{equation}
The same equality $\Phi_+=E$ continues to hold true as we go up in spin $j$ 
(we verified it for $j{=}\sfrac12$ and $j{=}1$), 
thus validating the  energy conservation $\pa^\mu T_{\mu 0}=0$.

\section*{Acknowledgements}
\noindent
K.K.~is grateful to Deutscher Akademischer Austauschdienst (DAAD) for the doctoral research grant~57381412. 
He thanks Gleb Zhilin for useful discussions. O.L.~benefitted from conversations with Harald Skarke.
Mathematica verification by Colin Becker and help with Figure~2 by Till Bargheer are acknowledged.


\end{document}